\documentclass{emulateapj}%{aastex}
\usepackage{url,graphicx,amssymb}
\usepackage{subfigure, lscape}
\long\def\symbolfootnote[#1]#2{\begingroup%
\def\thefootnote{\fnsymbol{footnote}}\footnote[#1]{#2}\endgroup}

\newcommand\swt{SWIFT J1753.5-0127}

\begin{document}

%%%%%%%%%%%%%%%%%%%%%%%%%%%%%%%%%%%%%%%%%%%%%%%%%%%%%%%%%%%%%%%%%%%%%%
%%%%      TITLE                                                   %%%%
%%%%%%%%%%%%%%%%%%%%%%%%%%%%%%%%%%%%%%%%%%%%%%%%%%%%%%%%%%%%%%%%%%%%%%
\shorttitle{Suzaku broadband spectroscopy of \swt.}
\shortauthors{Reynolds et al.}

\title
{Suzaku broadband spectroscopy of \swt~in the Low-Hard State}
\author{Mark T. Reynolds\altaffilmark{1}, Jon M. Miller\altaffilmark{1}, Jeroen
  Homan\altaffilmark{2} and Giovanni Miniutti\altaffilmark{3}}  
\email{markrey@umich.edu}

\altaffiltext{1}{Department of Astronomy, University of Michigan, 500 Church
  Street, Ann Arbor, MI 48109} 
\altaffiltext{2}{MIT Kavli Institute for Astrophysics and Space Research, 70
  Vasser Street, Cambridge, MA 02139} 
\altaffiltext{3}{LAEX, Centro de Astrobiologia (CSIC--INTA); LAEFF, P.O. Box
  78, E--28691, Villanueva de la Ca\~{n}ada, Madrid}

%%%%%%%%%%%%%%%%%%%%%%%%%%%%%%%%%%%%%%%%%%%%%%%%%%%%%%%%%%%%%%%%%%%%%%%%%%%%%
%%%%    Abstract                                                         %%%%
%%%%%%%%%%%%%%%%%%%%%%%%%%%%%%%%%%%%%%%%%%%%%%%%%%%%%%%%%%%%%%%%%%%%%%%%%%%%%

\begin{abstract}
We present \textit{Suzaku} observations of the Galactic black hole candidate
\swt~in the low-hard state. The broadband coverage of Suzaku enables us to
detect the source over the energy range 0.6 -- 250 keV.  The broadband
spectrum (2 -- 250 keV) is found to be consistent with a simple power-law
($\Gamma \sim$ 1.63). In agreement with previous observations of this
system, a significant excess of soft X-ray flux is detected consistent with
the presence of a cool accretion disc. Estimates of the disc inner radius
infer a value consistent with the ISCO ($\rm R_{in} \lesssim 6~R_g$, for
certain values of, e.g. $\rm N_H,~ i$), although we cannot conclusively rule
out the presence of an accretion disc truncated at larger radii ($\rm R_{in}
\sim 10 - 50~R_g$). A weak, relativistically-broadened iron line is also
detected, in addition to disc reflection at higher energy. However, the
iron-K line profile favours an inner radius larger than the ISCO (R$\rm _{in}
\sim 10 - 20~R_g$). The implications of these observations for models of the
accretion flow in the low-hard state are discussed.
\end{abstract}

\keywords{accretion, accretion discs - black hole physics - stars: binaries
  (Swift J1753.5-1027) X-rays: binaries} 

\maketitle
%%%%%%%%%%%%%%%%%%%%%%%%%%%%%%%%%%%%%%%%%%%%%%%%%%%%%%%%%%%%%%%%%%%%%%%%%%%%%
%%%%    Introduction                                                     %%%%
%%%%%%%%%%%%%%%%%%%%%%%%%%%%%%%%%%%%%%%%%%%%%%%%%%%%%%%%%%%%%%%%%%%%%%%%%%%%%
\section{Introduction}
The study of accretion processes is of fundamental astrophysical importance,
being ubiquitous on both the largest and smallest scales, i.e. ranging from
accretion by supermassive black holes in the center of galaxies at one end
to star and planet formation at the other.  Here we are interested in
studying accretion phenomena in the presence of a large gravitational
potential such as that observed in X-ray binaries (XRBs) and active galactic
nuclei (AGN). In these sources, the fundamental timescales of interest,
governing the accretion process, scale with the mass of the central object
($\rm \propto M_x$). Hence, studies of the Galactic XRB population provide
an excellent laboratory for detailed examination of the process of accretion
on humanly accessible timescales.

The very-high, high-soft and low-hard states (hereafter VHS, HSS \& LHS) are
the primary active accretion states observed in XRBs (See the review by
\citealt{I2} for a detailed description of accretion states in black hole
binaries). Despite being the most common mode of accretion in black hole
X-ray binaries, the nature of the accretion flow in the low-hard state
remains uncertain. Emission in the LHS is characterized by a hard power-law
spectrum ($\Gamma \sim$ 1.4 -- 1.7) and strong X-ray variability (30\% --
40\% RMS). Correlated variability at radio/NIR/optical wavelengths has also
been observed from the black hole systems while in the hard state
\citep{a15}. The hard power-law component may be the result of
Comptonization by a hot optically thin plasma of soft seed photons from a
thermal disc or magnetic structures (through cyclo-synchrotron processes -
see, e.g., \citealt{a35}). However, the geometry of this plasma is not well
understood. A popular model for the accretion geometry in the low-hard state
was given by \citet{a36}. In that model the standard thin accretion disc,
that dominates in the spectrally soft states is radially truncated and
replaced by an advection dominated accretion flow (ADAF: see
\citealt{a16}). The fundamental assumption of a radially recessed accretion
disc may find some support in the low disc reflection fractions which are
sometimes measured in the hard state (e.g., \citealt{a37}).

A number of models have been developed which do not require a
recessed disc in the low-hard state. \citet{a38} proposed that black hole
states are driven by the height and bulk velocity of magnetic flares above a
disc, which remains at the innermost stable circular orbit (ISCO). These
flares would serve to feed a mildly relativistically outflowing corona. Low
disc reflection fractions do not signal a recessed disc in this model, but
result from mild beaming of the hard X-ray flux away from the disc. A
similar model for the LHS, based on magnetically dominated coronae, has been
proposed by \citet{a39}. These outflowing coronae share some properties with
jet based models for the hard component \citep{a40}, in the sense that the
latter also produce low reflection fractions \citep{a25} without the need
for a recessed accretion disc.

A number of recent observations call into question the assumption of a
recessed disc in the LHS. In particular GX 339-4 was observed during its
2004 outburst at a luminosity of $\rm L \sim 0.05~L_{Edd}$ by both {\it
  RXTE} \& {\it XMM} \citep{a20}. The observations by {\it XMM} are critical
here as they provide coverage at energies below 3 keV whereas {\it RXTE} is
limited to energies above this. It was found that a cool disc blackbody
($\sim$ 0.35 keV) consistent with an optically thick geometrically thin
accretion disc extending to the innermost stable circular orbit (ISCO)
existed, in contrast to theoretical expectations. Fits with reflection
models revealed reflection fractions $\sim$ 0.2 -- 0.3. Previous
observations of Swift J1753.5-0127 yielded similar results for the cool disc
component, in this case a disc temperature of $\sim$ 0.2 keV was required
\citet{a18}; however, no significant disc reflection features were detected.

\begin{figure}
\begin{center}
\includegraphics[height=0.33\textheight,width=0.29\textwidth,angle=-90]{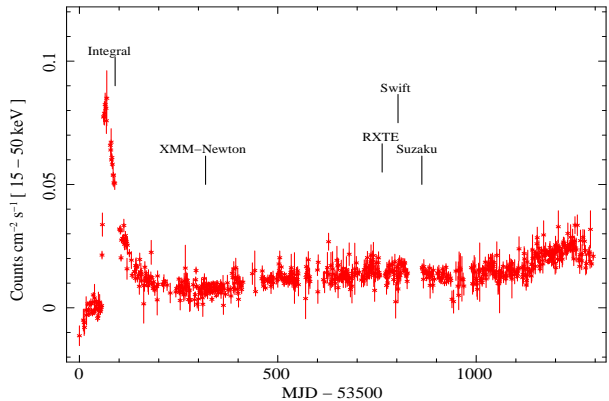}
\caption{{\it Swift} BAT hard X-ray lightcurve for \swt~from May 2005 to
  November 2008. The times of the {\it Integral} \citep{a19}, {\it XMM}
  \citep{a18}, {\it RXTE} \citep{a76}, \textit{Swift} (see \S 4.1) and
  \textit{Suzaku} (this paper) observations are indicted.}
\label{asmlc}
\end{center}
\end{figure}

\swt~was discovered by the \textit{Swift} burst alert telescope (BAT) at
X-ray and $\gamma$-ray energies on 2005 May 30 (\citealt{a1}). Subsequent
observations with the X-ray telescope (XRT) revealed a hard power-law
spectrum \citep{a2,a3}, while pointed \textit{RXTE} observations detected
0.6 Hz quasi periodic oscillations (QPO; \citealt{a7}). The system was also
detected at UV \citep{a8} and optical wavelengths \citep{a4}. Radio
observations with the \textit{MERLIN} array also detected a variable
counterpart \citep{a5}. Observations at optical wavelengths by
\citet{a69} have also detected a significant modulation with a period of
3.2hrs, which they identify as a superhump period slightly larger than the
actual orbital period. This would make \swt~the black
hole binary with the shortest known orbital period.

In addition to the observations of \citet{a18} above, a number of other high
energy studies of this source have been published, which we summarize below
(see Fig. \ref{asmlc}). An analysis of \textit{RXTE} observations of the
outburst was reported in \citet{a6,a17}. The X-ray spectrum was found to be
consistent with a power-law ($1.6 \leq \Gamma \leq 1.8$), while
low-frequency quasi-periodic oscillations (QPOs) were also detected (up to
0.9 Hz).  \citet{a19} analysed simultaneous \textit{RXTE} \&
\textit{INTEGRAL} data, which were also obtained during the 2005
outburst. The combined spectrum (3 -- 400 keV) could be fit with a model
consisting of thermal Comptonization modified by disc reflection ($\rm kT_0
\sim 0.5~keV, kT_e \sim 150~keV, \tau \sim 1, f \equiv \Omega/2\pi \sim
0.3$). QPOs were also detected here, but at a lower frequency than during
the outburst peak (0.24 Hz). 

To place meaningful constraints on the accretion flow in the LHS one
requires sensitivity to both the soft X-rays ($<$ 2 keV), in order to detect
the cool accretion disc, and higher energies in order to detect the most
prominent disc reflection features ( 5 -- 7 keV and 20 -- 30
keV). \textit{Suzaku} with its large bandpass and low background, is ideally
equipped to carry out these observations.  In this paper, we describe
observations undertaken with the \textit{Suzaku} X-ray observatory in 2007,
while \swt~was in the low-hard state.  In \S2, we describe our
observation and extraction of source spectra and lightcurves. We proceed to
analyze the data in \S3, where both phenomenological and more physically
motivated models are considered. The broadband spectrum (2 -- 250 keV) is
consistent with a simple power-law model, although there are also 
contributions from the accretion disc. In \S4, these results are discussed
in the context of models for the accretion flow in the low-hard state, and
finally our conclusions are presented in \S5.

%%%%%%%%%%%%%%%%%%%%%%%%%%%%%%%%%%%%%%%%%%%%%%%%%%%%%%%%%%%%%%%%%%%%%%%%%%%%%
%%%%    Observations                                                     %%%%
%%%%%%%%%%%%%%%%%%%%%%%%%%%%%%%%%%%%%%%%%%%%%%%%%%%%%%%%%%%%%%%%%%%%%%%%%%%%%
\section{Observations}
\swt~was observed while still in the low-hard state by \textit{Suzaku}
\citep{a9} from 2007 September 19 20:36 UT until September 22 10:30 UT
(obsid:402088010, PI: Homan, see Fig \ref{asmlc}). Data were acquired over a
broad spectral range (0.2 -- 600 keV), with the X-ray imaging spectrometer
(XIS: \citealt{a10}) and the hard X-ray detector (HXD: \citealt{a11,a12}).
The source was observed at the XIS nominal position for total uncorrected
exposure times of $\sim$ 95 ks \& 82 ks respectively .  

All data reduction and analysis takes place within the \textsc{heasoft
6.4.1} environment, which includes \textsc{ftools 6.4, suzaku 8.0} and
\textsc{xspec 12.4.0x}. The latest versions of the relevant \textit{Suzaku}
\textsc{caldb} files were also used.

\subsection{X-ray Imaging Spectrometer}
The XIS is installed at the focal plane of the four X-ray telescopes (XRT:
\citealt{a13}) and currently consists of 3 functioning detectors XIS0, XIS1
and XIS3. XIS0 \& XIS3 are front illuminated and provide coverage over the
energy range 0.4 - 12 keV whereas XIS1 is back illuminated in an effort to
provide greater sensitivity at lower energies, 0.2 -- 12 keV. 
The XIS has a field of view of $\sim$ 18' x 18' (1024$^2$ pixels) and was
operated in 5x5, 3x3 readout mode. In addition the data were taken in 1/4
window mode in an effort to minimize possible photon pile-up, giving a time
resolution of 2s. 

As the downloaded data products had been processed via the \textit{Suzaku}
pipeline v2.1.6.15, the data were reprocessed from the raw telemetry files
as recommended in the \textit{Suzaku} data reduction guide (abc guide)
\footnote{http://heasarc.nasa.gov/docs/suzaku/aehp\_data\_analysis.html}.
Standard screening was applied, in particular, we extracted \textit{ASCA}
event grades 0:0, 2:4, 6:6 with the data filtered to be taken outside of the
South Atlantic Anomaly (SAA) and where the earth elevation angle was greater
than 5$^{\circ}$. Good time interval (GTI) events were extracted using
\textsc{xselect}, where the 3x3 and 5x5 observation mode data for each
detector were extracted simultaneously. Science images, spectra and
backgrounds were then extracted from these event files.

Even though the observations were carried out in 1/4 window mode, we
nonetheless suffered from pileup at the source position. Hence the spectra
were extracted using an annular extraction region extending from 30 to 250
pixels from the source position. This extraction region size was chosen so
as to extract $>$ 99$\%$ of the point source flux.  As the outer radius of
this annulus is larger than the window size ($\sim$ 280, 295, 280 pixels
respectively), the effective extraction region is the intersection of window
and the annulus. The resulting extraction region has an area equivalent to
62\% of the 250 pixels outer radius, hence, we expect to have detected a
commensurate percentage of the total source flux. 

Background spectra were extracted from a source free region of the detector,
these are automatically scaled to match the data during the spectral
analysis.  Response files were generated using the tasks \textsc{xisrmfgen}
and \textsc{xissimarfgen}.  The background and response files were then
grouped with the science spectrum for analysis in \textsc{xspec}.

\begin{figure}
\begin{center}
\includegraphics[height=0.33\textheight,angle=-90]{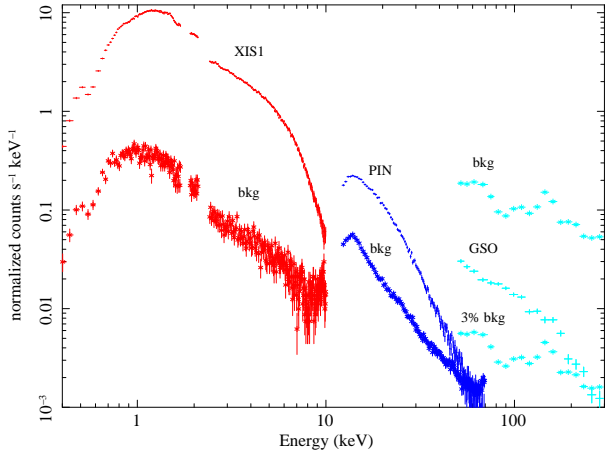}
\caption{Background subtracted XIS1, PIN and GSO spectra and their
  associated background spectra. For clarity, we plot only a
  single XIS detector. \swt~is detected by the PIN detector
  out to 60 keV, while the GSO detects flux to $\sim$ 250 keV assuming the
  background is reproducible to an accuracy of 3\% (see text).}  
\label{xispingso_back}
\end{center}
\end{figure}

\subsection{Hard X-Ray Detector}
The HXD covers the energy range from 10 -- 600 keV, consisting of two
separate detectors, (i) PIN: Silicon PIN photodiodes covering the energy
range 10 -- 70 keV and, (ii) GSO: GSO/BGO phoswich scintillators covering
the energy range 40 - 600 keV. Due to the arrangement of the instrument,
with the PIN diodes residing in front of the GSO scintillator in each of the
16 detector units that make up the HXD, the raw data do not differentiate
between the PIN \& GSO, with this distinction instead being made during the
extraction process.

As at the time of writing, the GSO analysis procedure had not been included in
the official \textit{Suzaku} data reduction pipeline, all of the hard X-ray
detector data were reprocessed following the prescription in the abc guide
and the 7-step
guide\footnote{astro.isas.ac.jp/suzaku/analysis/7step\_HXD\_20080114.txt}.
GTI science events were extracted using \textsc{xselect} with the
appropriate GTI files and filter options i.e., pointing elevation $>$
5$^{\circ}$ above earth and excluding data taken near the SAA. The relevant
events for each detector were then extracted by requiring DET\_TYPE=1:1 and
DET\_TYPE=0:0 for the PIN and GSO respectively.

The relevant background and response (pin/gso *20080129.rsp) files were
obtained from the \textit{Suzaku} website. A GTI between the data and the
background file was then created using the ftool \textsc{mgtime}, which was
then used to to extract the spectra in \textsc{xselect}. Standard
corrections were applied to the data, i.e. deadtime for science
data, PIN background exposure time. As the HXD background files do not
include the contribution from the cosmic X-ray background (CXB), the
expected CXB was simulated following the recipe detailed in the abc guide,
this was then added to the background file. As a check on the accuracy of
the background, a separate background estimate was made using the earth
occulted data (earth elevation = 0). This was found to be consistent with
the above background.

A similar procedure is followed in the case of the GSO spectral extraction
with the following caveat: the background file exposure time does not need
to be corrected, instead one must rebin the science spectrum to match the
provided background. The sensitivity of the GSO detector is background
dominated, hence the high energy detection threshold is determined not by
the statistical error but by the reproducibility of the background, here we
conservatively estimate the background reproducibility to be 3\% (e.g
\citealt{a53}). In Fig. \ref{xispingso_back}, we plot the extracted science
\& background spectra. \swt~is detected out to an energy of 250 keV

%%%%%%%%%%%%%%%%%%%%%%%%%%%%%%%%%%%%%%%%%%%%%%%%%%%%%%%%%%%%%%%%%%%%%%%%%%%%%
%%%%      Analysis & Results                                             %%%%
%%%%%%%%%%%%%%%%%%%%%%%%%%%%%%%%%%%%%%%%%%%%%%%%%%%%%%%%%%%%%%%%%%%%%%%%%%%%%
\section{Analysis \& Results}

\subsection{Light Curves}
Source and background lightcurves were extracted from the event files using
\textsc{xselect} and the appropriate good-time-interval events, after the
application of the appropriate baryocentric correction using the ftool
\textsc{aebarycen}. The flux from \swt~is observed to remain constant
throughout our observation. The mean count rates for each individual
detector are approximately : 16.2, 20.8, 17.1, 2.6, 2.2 counts
s$^{-1}$(XIS0, XIS1, XIS3, PIN, GSO).

There is no evidence for any periodic modulation, the lightcurves across all
detectors are characterized by rapid variability of a stochastic nature as
is expected from accretion processes in the vicinity of a black hole
\citep{a50}. This is consistent with RXTE power spectra bracketing this data
that revealed a power-law slope, $\beta \sim -1$.

\begin{figure*}
\begin{center}
\subfigure[$\rm N_H = 0.18 \times 10^{22}~cm^{-2},~i \sim 63$]{\includegraphics[height=0.35\textheight,angle=-90]{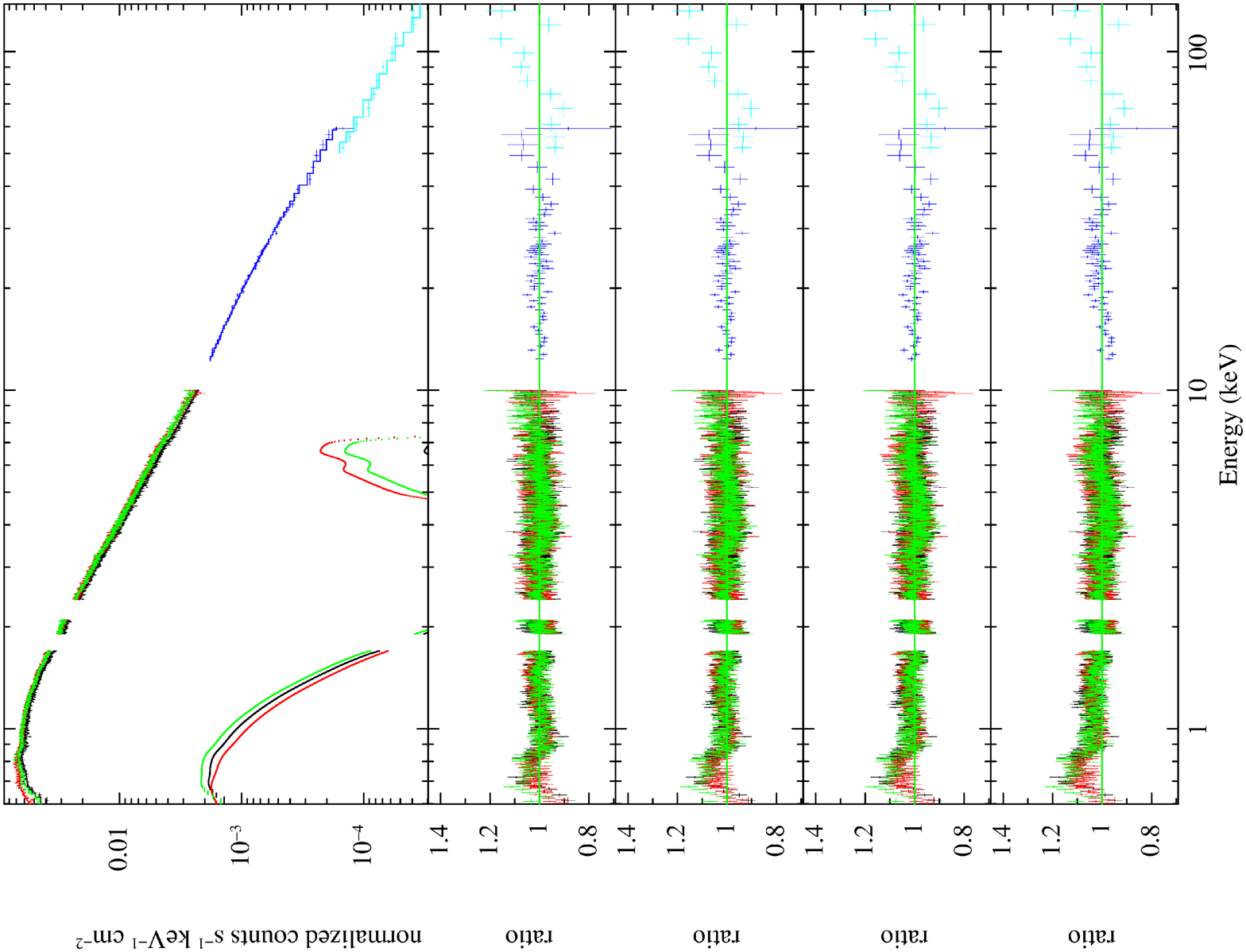}}
\subfigure[$\rm N_H = 0.23 \times 10^{22}~cm^{-2},~i \sim 63$]{\includegraphics[height=0.35\textheight,angle=-90]{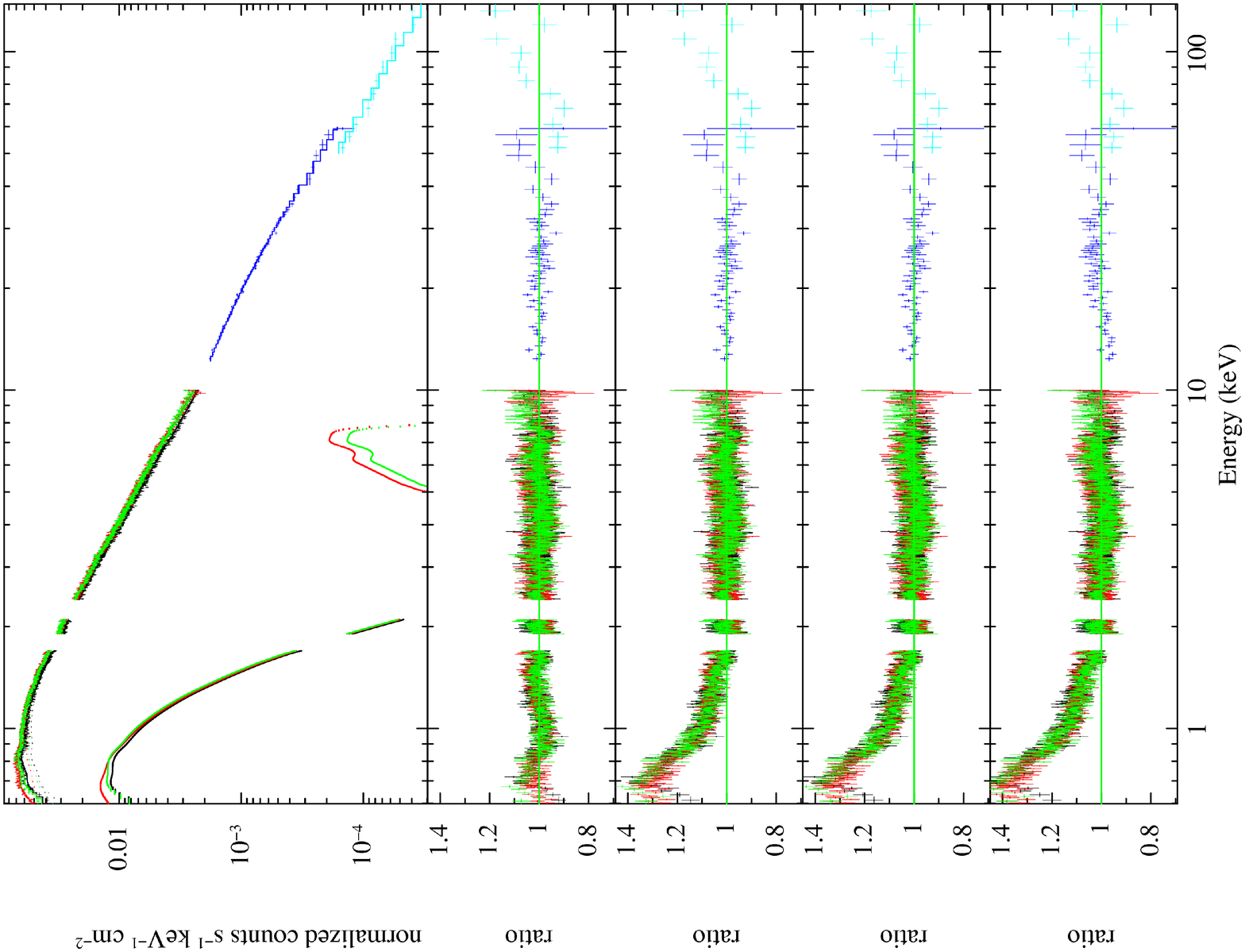}}
\caption{Best fit to the \textit{Suzaku} spectra of SWIFT J1753.5-0127 in
  the 0.6 - 150 keV range. The best fit model consisting of a disc blackbody
  plus power-law modified by disc reflection ({\tt
    pha*(reflect*(diskbb+po)+laor)}) is plotted. The residuals show from
  bottom to top {\tt pha*po}, {\tt pha*(reflect*po)}, {\tt
    pha*(reflect*(po)+laor)}, {\tt pha*(reflect*(diskbb+po)+laor)}. There is
  a clear excess at soft X-ray energies, the solid lines denote the {\tt
    diskbb} \& {\tt laor} components. The regions 1.7--1.9 keV and 2.1--2.4
  keV are ignored due to the presence of instrumental features.}
\label{soft_excess}
\end{center}
\end{figure*}

\subsection{Spectra I: Phenomenological Models}\label{spec_i}
We initially choose to fit the spectra with a number of phenomenological
models in an effort to provide a broad characterization of the data. Initial
spectral fits were made to the entire dataset consisting of 5 spectra in
total (XIS0, XIS1, XIS3, PIN, GSO), spanning the energy range 0.6 -- 250 keV
over which SWIFT J1753.5-0127 is reliably detected. Explicitly, the spectra
provide us with data in the following ranges: XIS0, XIS3 0.4 -- 10 keV; XIS1
0.2 -- 10 keV; PIN 12 -- 70 keV and GSO 50 -- 250 keV. Unfortunately, the
low energy response of the XIS detectors contains a number of uncertain
residuals, hence we additionally ignored the regions below 0.6 keV and
between 1.7 -- 1.9 keV and 2.1 -- 2.4 keV in all further modelling.

Initially, the column density was fixed at a value consistent with previous
detections ($\rm N_H = 2.3 \times 10^{21}~cm^{-2}$: \citealt{a18}), while
the normalization was allowed to vary independently for each spectrum. The
resulting normalizations were found to be consistent with those expected for
the \textit{Suzaku} detectors. The spectra were fit with a model consisting
of an absorbed power-law ({\tt pha*po})\footnote{Throughout this paper the
  abundances and cross-sections assumed are {\tt bcmc} \citep{a78} and
  {\tt angr} \citep{a79} respectively}, which provided an acceptable fit
except at the lowest energies ($\leq$ 2 keV).  As the main residual was
present at energies below 2 keV, this region was ignored in further
fitting. Re-applying the power-law fit results in a value for the spectral
index of $\rm \Gamma = 1.62$ ($\chi^2/\nu = 6968/6463$).  It is immediately
apparent that a simple phenomenological power-law model suffices to describe
the spectrum from 2 -- 250 keV.

The data was also fit with a {\tt cutoffpl} model for comparison. The fit
does not indicate the presence of a cut-off in the spectrum with the {\tt
  cutoffpl} model fit being significantly inferior to a simple power-law,
i.e we find $\chi^2/\nu_{po}$ = 6898/6463 (6968/6456) and
$\chi^2/\nu_{cutoffpl}$ = 6962/6462 (7048/6455) for the $\rm N_H =
0.18~(0.23) \times 10^{21}~cm^{-2}$ models respectively. In both cases the
best fit cut-off power-law model requires the high energy cut-off to be 500
keV, i.e the intrinsic upper limit of the {\tt cutoffpl} model. Ignoring the
data above 150 keV does not improve the quality of the {\tt cutoffpl} fit
relative to the {\tt po} fit.

As the GSO background contains a large feature in the energy range 150 --
180 keV (see Fig. \ref{xispingso_back}), we decided to ignore the data
beyond 150 keV in all further fitting. Furthermore, as there is clearly an
additional component contributing at energies below 2 keV, this region is
ignored while constraints are placed on the hard X-ray emission. Repeating
the above power-law fit, we find $\rm \Gamma = 1.61$ ($\chi^2/\nu =
6869/6456$)\footnote{All further fits in this section assume $\rm N_H =
0.18 \times 10^{21}~cm^{-2}$ unless otherwise explicitly stated}.

Inspection of the residuals reveals any contribution from disc reflection to
be small, although there is evidence for curvature in the PIN spectrum. To
investigate the possible contribution due to disc reflection, the best fit
power-law from above was convolved with the reflection model of \citet{a43}
({\tt reflect*po}).  The inclination was held fixed (cos{\it i} = 0.45),
while the abundances of metals were frozen at the default values ($\rm
N_Z/N_{\sun}$ = 1). The best fit model reveals a slightly softer power-law
($\rm \Gamma \sim 1.63$) in addition to a highly significant ($> 9\sigma$ as
determined via an F-test) reflection fraction, $\rm f \sim 0.26$
($\chi^2/\nu = 6778/6455$). As the measured spectral parameters depend on
the inclination angle, these and subsequent fits were repeated at a lower
inclination angle of 30 degrees. These fits revealed a lower reflection
fraction, f $\sim$ 0.14 (see Table \ref{specfit_params}).

Closer inspection of the XIS spectra residuals reveals a slight excess
consistent with the presence of a broad Fe K line. To place a constraint on
the size of any possible line, a Gaussian was added to this model. The best
energy of this line is $\sim$ 6.4 $\pm$ 0.1 keV, in agreement with that
expected from neutral Fe K$\alpha$. This is consistent with that expected
from reflection from neutral matter ($\rm E_{line} = 6.4 keV$) as assumed in
the {\tt reflect} model. The Gaussian component was then replaced with
relativistic line model ({\tt laor}: \citealt{a45}), which more accurately
represents the expected line profile in the inner disc region. The
inclination of this line is fixed at the same value as the the reflection
component, the emissivity profile of the disc is fixed at $\rm R^{-3}$ and
the outer radius of the emitting region is fixed at 400 R$\rm _g$. We find
the data require the presence a broad weak iron line ($\rm > 8\sigma$
significant as determined via an F-test; EW = 73$\rm \pm30~eV$). Allowing
the inner disc radius to vary, we find $\rm R_{in} = 19^{+6}_{-4}~R_g$. The
iron line inner radius has a strong inclination dependence with a best fit
$\rm R_{in} \sim 13~R_g$ at 30$^{\degr}$. The above inner radii are
inconsistent with the ISCO for a Schwarzschild black hole, if we extend the
confidence intervals we place the following limits on the inner radius --
$\rm R_{in} \geq 11~R_g~\&~6.8~R_g$ (3$\sigma$ level) for inclinations of
63.256$^{\degr}$ \& 30$^{\degr}$ (cos{\it i} = 0.45, 0.8660) respectively.
 
Including the lower energy flux in the fit once again, results in a
chi-squared value of $\chi^2/\nu = 9431/7450$, we plot this in
Fig. \ref{soft_excess}.  A soft excess is present consistent with previous
observations, i.e. \citet{a18}. A simple blackbody accretion disc component
({\tt diskbb}: \citealt{a34}) was added to this model to account for the
excess soft X-ray flux, while the power-law index, $\Gamma$, and the
reflection fraction, f, were frozen at their previous best fit values. The
disc component is strongly required by the data; the best fit is achieved
for a disc temperature of kT = 0.20 keV ($\chi^2/\nu = 8149/7442$). We
measure the associated 0.6 -- 10 keV (2 -- 150 keV) unabsorbed flux to be
$\rm 6.8 \times 10^{-10}~erg~s^{-1}~cm^{-2}$ ($\rm 2.4 \times
10^{-9}~erg~s^{-1}~cm^{-2}$). 
 
The column density is crucial here: although it has little impact on the
spectrum above energies $\sim$ 2 keV, it will have a significant impact on
the shape of the spectrum and the measured flux below this value,
e.g. Fig. \ref{soft_excess}.  In order to test the effect of different
values for $\rm N_H$ the fitting was repeated at a number of different
values for the interstellar column density ranging from $\rm N_H = 0.17
\times 10^{22}~cm^{-2}$ \citep{a30} to $\rm 0.28 \times 10^{22}~cm^{-2}$
\citep{a29}. We find the best fit as measured by \textit{Suzaku} was found
to be $\rm N_H = (0.18\pm 0.01) \times 10^{22}~cm^{-2}$. In Table
\ref{specfit_params}, we list the parameters for the best fit model and also
those for the model corresponding to the value of $\rm N_H$ measured
previously by \textit{XMM}.

\begin{figure}
\begin{center}
\includegraphics[height=0.33\textheight,angle=-90]{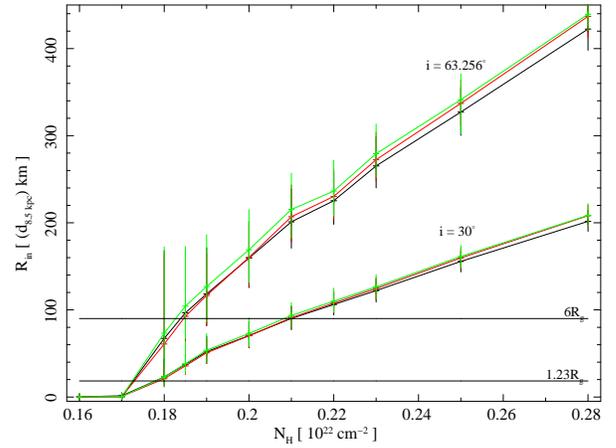} 
\caption{Inner disc radius for the cool multi-colour blackbody accretion
  disc {\tt diskbb} component versus column density, where we have corrected
  for spectral hardening and the inclination (see text). The two sets of lines
  indicate the expected inner radius assuming an inclination of 30$^{\degr}$
  \& 63.256$^{\degr}$ respectively.  The error bars correspond to the 90\%
  confidence interval for the disc normalization. The horizontal lines
  indicate the position of the ISCO for both a Kerr (1.23R$\rm_g$) \&
  Schwarzschild (6R$\rm_g$) black hole, where the black hole mass is assumed
  to be 10$\rm M_{\sun}$ ($\rm R_g \equiv GM/c^2$).}
\label{inner_radius}
\end{center}
\end{figure}

The normalization of the {\tt diskbb} model is proportional to the inner
disc radius, norm = ($\rm r_{in}$ [km])/d [10kpc])$^2$cos$\theta$. In
Fig. \ref{inner_radius}, we plot the inner disc radius corresponding to the
best fit {\tt diskbb} component for various values of the column density,
where we have corrected for spectral hardening and the inner disc radius
(see \S \ref{thin_disc_isco}). We find that for reasonable values of the
column density and inclination the inner disc radius may reside close to the
ISCO, in agreement with recent work, which has provided evidence that the
cool disc may reside at or near the ISCO, even at the low luminosities
typically observed in the LHS \citep{a18,a20, a32}. We also modelled the
soft excess using the {\tt ezdiskbb} and {\tt diskpn} models in
\textsc{xspec}. In both cases, the observed blackbody component is found to
be consistent with that obtained using {\tt diskbb}. We also experimented
with using the {\tt kdblur} kernel to relativistically blur the above
continuum model; however, this did not result in an improved fit.

We also carried out fits to the individual telemetry segments of the XIS,
PIN \& GSO spectra, to check for possible variability. The fits to each of
the 3 segments were found to be consistent with each other, hence there is
no evidence for spectral variability in agreement with the constant flux
observed from the lightcurves.

%\begin{landscape}
\begin{table*}
\begin{center}
\caption{Spectral Fit Parameters I\label{specfit_params}}
\begin{tabular}{lccccc}
\tableline
 Model & Parameter & \multicolumn{4}{c}{Value} \\[0.5ex]
\tableline
{\tt phabs } & $\rm N_H~[ 10^{22}~cm^{-2} ]$ & \multicolumn{2}{c}{0.23} & \multicolumn{2}{c}{0.18}\\[0.5ex]  
             & $\rm i$            & 30     & 63.256 & 30 & 63.256\\[0.5ex]
\tableline
{\tt po}     & $\rm \Gamma$ & 1.619$\pm$0.003 & 1.619$\pm$0.003 &
1.608$\pm$0.003 & 1.608$\pm$0.003 \\[0.5ex]
$\chi^2$ \hspace{5mm}($\nu = 6456$)&  & 6938 & 6938 & 6869 & 6869 \\[0.5ex]
\tableline
{\tt reflect*po} & $\rm \Gamma$            & 1.640$\pm$0.005&
1.649$\pm$0.006 & 1.625$\pm$0.005 & 1.632$\pm$0.005\\[0.5ex]
                 & $\rm f~[ \Omega/2\pi ]$ & 0.17$\pm$0.03  & 0.33$\pm$0.05
& 0.14$\pm$0.03 & 0.26$\pm$0.05 \\[0.5ex]
$\chi^2$ \hspace{5mm}($\nu = 6455$)    &  & 6830 & 6807 & 6793 & 6778 \\[0.5ex]
\tableline
{\tt reflect*po+laor} & $\rm \Gamma$             & 1.645$\pm$0.005 &
1.654$\pm$0.005 & 1.630$\pm$0.005 & 1.637$\pm$0.005\\[0.5ex]
                      & $\rm f~[ \Omega/2\pi ]$  & 0.15$^{+0.03}_{-0.02}$ &
0.27$\pm$0.05 & 0.12$\pm$0.03 & 0.21$\pm$0.05\\[0.5ex]
                      & $\rm E_{line}~[ keV ]$    & 6.4$\pm$0.1 &
6.4$\pm$0.1 & 6.4$\pm$0.1  & 6.4$\pm$0.1\\[0.5ex]
                      & $\rm R_{in}~[ R_g ] $ & 13$^{+7}_{-4}$     & 19$^{+4}_{-5}$
& 13$^{+7}_{-5}$ & 19$^{+6}_{-4}$\\[0.5ex]
                      & $\rm EW~[ eV ] $ & 60$\pm$30  & 80$^{+50}_{-40}$
& 60$^{+20}_{-35}$ & 70$\pm$30\\[0.5ex]
$\chi^2$ \hspace{5mm}($\nu = 6451$)    & & 6743 & 6720 & 6707 & 6694\\[0.5ex]
\tableline
{\tt reflect*(diskbb+po)+laor} & $\rm T_{diskbb}~[ keV ]$  & 0.190$\pm$0.002
& 0.180$\pm$0.003 & 0.25$\pm$0.03& 0.20$\pm$0.04 \\[0.5ex]
                               & $\rm norm0_{diskbb}$ & 3866$\pm$420 &
4830$^{+620}_{-470}$ & 137$^{+124}_{-63}$ & 361$^{+1052}_{-232}$  \\[0.5ex]
                               & $\rm norm1_{diskbb}$ & 4023$\pm$410 &
5080$^{+600}_{-460}$ & 115$^{+97}_{51}$ & 294$^{+780}_{-183}$\\[0.5ex]
                               & $\rm norm3_{diskbb}$ & 4144$\pm$460 &
5340$^{+670}_{-530}$ & 135$^{+136}_{-66}$ 76 & 432$^{+1536}_{-293}$\\[0.5ex]
$\chi^2$ \hspace{5mm}($\nu = 7442$)   & & 8349  & 8282 & 8160 & 8149 \\[0.5ex]
\tableline
\end{tabular}
\tablecomments{Results of fits to the \textit{Suzaku} spectra for
  \swt, spanning the energy range 0.6 -- 150 keV. All models are modified by
  ``{\tt pha}'' to account for interstellar extinction, which was frozen 
  at the value indicated in the table above. In addition, the spectral
  regions below 0.6 keV and between 1.7 -- 1.9 keV and 2.1 -- 2.4 keV are
  ignored at all times due the presence of instrumental calibration
  uncertainties. All errors are quoted at the 90\% confidence level.}
\end{center}
\end{table*}
%\end{landscape}

\subsection{Spectra II: Physically Motivated Models}\label{spec_ii}
While the above models provide an excellent fit to the observed data, due to
their phenomenological nature they offer limited constraints on the nature
of the accretion flow. In this section, we will consider more complex models
in an effort to place improved constraints on the physical processes that
create the observed spectrum. We consider two scenarios for the observed
spectrum, (i) the observed spectrum is the result of the reflection of hard
X-rays from a power-law incident on the accretion disc , and (ii) the
spectrum is due to emission from a Comptonizing corona.\\

\noindent i) Disc Reflection Models:\\ 
Observations of a number of AGN and Galactic black holes have revealed disc
reflection features, the most prominent of which are the Fe K line at $\sim$
6.4 keV and the reflection bump at $\sim$ 20 -- 30 keV.  The constant
density ionized disc model ({\tt CDID}: \citealt{a21,a22}) is used to
model this effect. Here, we only consider data below 60 keV as the {\tt
  CDID} model as implemented in \textsc{xspec} is not valid at energies
above 100 keV. Solar metallicity is assumed in these fits.

\begin{figure}
\begin{center}
\includegraphics[height=0.33\textheight,angle=-90]{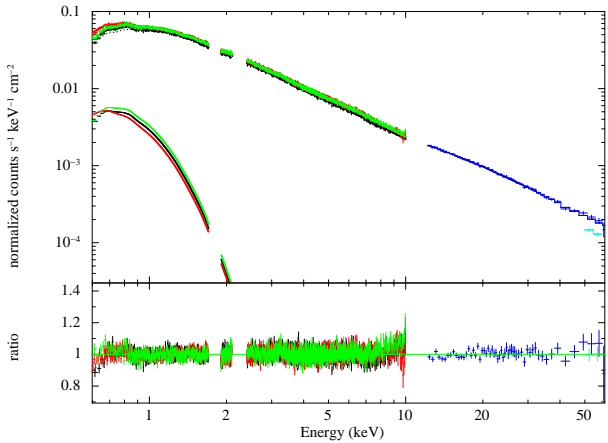}
\caption{The best fit disc reflection model ({\tt pha*(diskbb+CDID)}) to the
  \swt~spectrum, see \S\ref{spec_ii} for details. The inferred inner disc
  radius is low and consistent with that measured in \S\ref{spec_i}. See
  Table \ref{specfit_params1} for the model parameters.}
\label{spec_ii_images_1}
\end{center}
\end{figure}

As a check a power-law fit to the 2 -- 60 keV region of the spectrum alone
was carried out, the resulting power-law index is found to be in agreement
with the value derived from a fit to the entire spectrum (\S \ref{spec_i}).
Initially the spectrum was fit with the {\tt CDID} model alone modified by
interstellar absorption, the resulting fit was good ($\chi^2/\nu$ =
8070/7428), with a moderately ionized disc ($\xi \sim$ 3.55 ) and a
reflection fraction ($\rm f \sim 0.14$) in agreement with that measured 
earlier. A blackbody accretion disc component was added to this model
significantly improving the fit ($\chi^2/\nu$ = 8017/7424, $\rm >~6\sigma$
significant as measured by an ftest).  For our best fit model {\tt
  pha*(diskbb+CDID)}, we find the following parameters $\rm N_H \sim 0.20,
T_{ddiskbb} \sim 0.19, \xi \sim 3.55, \Gamma \sim 1.61, f \sim 0.12$ , see
Table \ref{specfit_params1} \& Fig \ref{spec_ii_images_1}.

The disc is found to be moderately ionized, while the low reflection
fraction is consistent with the absence of any large reflection features in
the observed spectrum. The disc temperature in this model is consistent with
that found assuming a simple {\tt diskbb+po} model (see \S\ref{spec_i}),
$\rm T_{diskbb} \sim 0.19~keV$.  The inner radius in this case is $\rm \sim
70~cos\theta^{-1}~d_{8.5kpc}~km$. We note that for inclinations less than
39$^{\degr}$ this is less than the radius of the ISCO for a 10 M$_{\sun}$
Schwarzschild black hole. \\

\noindent ii) Comptonizing Corona Models:\\ 
X-ray spectra of black hole binaries in the low-hard state are typically
modelled assuming the hard X-ray flux originates in a corona lying above the
accretion disc, which then scatters the soft X-ray flux from the disc to
higher energies.  \textit{Suzaku} observations of GRO J1655-40 \& Cyg X-1
have been modelled in such a manner \citep{a27,a28}, here we model the
broadband spectrum of \swt~following the Comptonization model outlined in the
above papers.

Initial fits with a single Comptonizing component were found to provide an
inadequate fit, particularly at the highest energies. As in the cases of
GRO J1655-40 and Cyg X-1 above, we model the spectrum using a pair of
Comptonizing coronae with the same electron temperature but differing
optical depths ({\tt diskbb+compps+compps+laor}), where a {\tt laor}
component is added to account for the possible presence of a relativistic
Iron line. A spherical geometry was assumed for the Comptonizing cloud and
possible disc reflection effects were accounted for using the reflection
routine built into the {\tt compps} model. Firstly the fit was carried out
without the accretion disc component (i.e. {\tt pha(compps+compps)}) to
check if this component was actually required, the best fit is good
($\chi^2/\nu$ = 8173/7430). However, the column density in this case is
large $\rm N_H$ ($\rm \sim 0.28 \times 10^{22}~cm^{-2}$), while the seed
temperature is low $\rm T_{in1} = T_{in2} = 0.1$. Significant disc
reflection is not required in this model.

Addition of an accretion disc to this model significantly improved the
quality of the fit, with a reduced chi-squared of 1.09 ($\chi^2/\nu$ =
8086/7427). The disc component is not required in the fit to the XIS0
spectrum; however, the 90\% upper limit to the disc normalization is
consistent with the value measured from the XIS1 \& XIS3 spectra. Again we
note that the value returned for the interstellar extinction $\rm N_H \sim
0.31 \times 10^{22}~cm^{-2}$ is much higher than any previously measured
value, and as such would appear to be inconsistent with expectations for
$\rm N_H$ in the direction of \swt~ (see \S \ref{nh_discuss}).  The iron
line component also significantly improved the fit ($\chi^2/\nu$ =
8009/7422; see Fig. \ref{spec_ii_images_2}), where the inner radius is
$\sim$ 60 R$\rm _g$. 

In the best fit model above the temperature of the electrons in the corona
is low $\rm kT_e \sim 53~keV$, while the optical depths are found to be
approximately 0.34 \& 2.57 for the 2 Comptonizing components respectively,
see Table \ref{specfit_params1}. We may estimate the size of the optically
thin ($\rm \tau \sim 0.34$) and optically thick ($\rm \tau \sim 2.57$)
regions of the Comptonizing cloud from the normalization, where the radius
of the spherical cloud is $\rm R \sim d_{10 kpc}\times(cos\theta\times
norm)^{0.5}$. In both cases the inferred radius is small $\sim$ 305$\rm
cos\theta^{0.5}$ km \& 540$\rm cos\theta^{0.5}$ km ($\rm \sim
20~R_g~\&~36~R_g$ assuming a 10 $\rm M_{\sun}$ black hole and a distance of
8.5 kpc) for the optically thick and optically thin regions respectively. The
temperature of the seed disc component is low at 0.1 keV. We note that this
is approximately half the value for the blackbody temperature one finds when
fitting the data assuming a simple reflection continuum (see \S\ref{spec_i})
or when fitting the data with a detailed reflection model (see
\S\ref{spec_ii}i). As the temperature is low and the lower limit to our data
is only 0.6 keV, the disc normalization is poorly constrained as may be seen
by inspection of Table \ref{specfit_params1}. Nonetheless, the inner radius
of the accretion disc is consistent with overlapping with the coronal region
(i.e. $\rm R_{in-disc} \approx R_{corona}$ within the errors). 

\begin{table}
\begin{center}
\caption{Spectral Fit Parameters II\label{specfit_params1}}
\begin{tabular}{lcc}
\tableline
 Model & Parameter & Value \\[0.5ex]
\tableline
{\tt phabs } & $\rm N_H~[ 10^{22}~cm^{-2} ]$ & 0.20$\pm$0.01\\[0.5ex]  
{\tt diskbb+cdid}     & $\rm T_{diskbb}~[keV]$ & 0.19$\pm$0.01  \\[0.5ex]
   & $\rm norm_{xis0}$ & 1554$\rm ^{+824}_{-783}$  \\[0.5ex]
   & $\rm norm_{xis1}$ & 1554$\rm ^{+1222}_{-670}$  \\[0.5ex]
   & $\rm norm_{xis3}$ & 1554$\rm ^{+1338}_{-886}$  \\[0.5ex]
   & $\rm \Gamma$ & 1.614$\pm$0.002  \\[0.5ex]
   & $\rm f~[\Omega/2\pi]$ & 0.12$\rm^{+0.02}_{-0.01}$  \\[0.5ex]
   & $\rm \xi~[erg~s^{-1}~cm^{-2}]$ & 3.47$\rm^{+0.08}_{-0.05}$  \\[0.5ex]
   & $\chi^2$ ($\nu = 7424$)& 8017  \\[0.5ex]
\tableline
             & Parameter  & Value \\[0.5ex]
\tableline
{\tt phabs } & $\rm N_H~[ 10^{22}~cm^{-2} ]$ & 0.31$\pm$0.01 \\[0.5ex]
{\tt diskbb+compps} & $\rm T_{diskbb}~[ keV ]$ & 0.10$\pm$0.01 \\[0.5ex]
{\tt +compps+laor}  & $\rm norm_{xis0}$ & $\leq 3.3\times 10^5$ \\[0.5ex]
  & $\rm norm_{xis1}$ & $\rm (0.35^{+1.7}_{-0.2})\times 10^{5}$ \\[0.5ex]
  & $\rm norm_{xis3}$ & $\rm (1.84 \pm 1.4)\times 10^{5}$ \\[0.5ex]
  & $\rm KT_e~[keV]$ & 53$\pm$3 \\[0.5ex]
  & $\rm T_{in1}~[keV]$ & $\rm T_{diskbb}$ \\[0.5ex]
  & $\rm \tau_1$       & 0.34$\pm$0.01 \\[0.5ex]
  & $\rm norm_{compps1}$ & $\rm (4.14^{+0.13}_{-0.21})\times 10^{5}$ \\[0.5ex]
  & $\rm T_{in2}~[keV]$ & $\rm T_{diskbb}$ \\[0.5ex]
  & $\rm \tau_2$       & 2.57$^{+0.01}_{-0.05}$ \\[0.5ex]
  & $\rm norm_{compps2}$ & $\rm (1.28^{+0.05}_{-0.02})\times 10^{5}$\\[0.5ex]
  & $\rm f [\Omega/2\pi]$ & $\leq 0.01$ \\[0.5ex]
  & $\rm E_{line}~[keV]$ & 6.4$\pm$0.1 \\[0.5ex]
  & $\rm R_{in}~[R_g]$ & 60$^{+70}_{-25}$ \\[0.5ex]
  & $\rm EW~[eV]$ & 70$\pm$30 \\[0.5ex]
  & $\chi^2$ ($\nu = 7422$)& 8009 \\[0.5ex]
\tableline
\end{tabular}
\tablecomments{Parameters for the physically motivated models from section
  \ref{spec_ii}. All errors have been calculated using the {\tt error}
  command in \textsc{xspec} and are quoted at the 90\% confidence level. For
  {\tt compps} models the inclination is frozen at 63.256$^{\degr}$, due to
  the spherical geometry of the corona the inclination dependence of this
  model is negligible.}
\end{center}
\end{table}

%%%%%%%%%%%%%%%%%%%%%%%%%%%%%%%%%%%%%%%%%%%%%%%%%%%%%%%%%%%%%%%%%%%%%%%%%%%%%
%%%%      Discussion                                                     %%%%
%%%%%%%%%%%%%%%%%%%%%%%%%%%%%%%%%%%%%%%%%%%%%%%%%%%%%%%%%%%%%%%%%%%%%%%%%%%%%
\section{Discussion}
We present \textit{Suzaku} broadband spectra (0.6 -- 250 keV) of the black
hole candidate \swt~while in the LHS. The observed spectrum is measured to
be consistent with an unbroken power-law, $\Gamma \sim 1.63$. During our
observation the flux from the source was observed to be constant, with a 0.6
- 150 keV unabsorbed flux of $\rm 2.6 \times 10^{-9}~erg~s^{-1}~cm^{-2}$
($\rm L_x/L_{Edd} = 0.016~d_{8.5kpc}^2M_{10M_{\sun}}$). This is consistent
with \textit{INTEGRAL} observations of GRO J1655-40 in the LHS where
unbroken power-law emission ($\Gamma \sim 1.72$) extending out to $\sim$ 500
keV was detected at a luminosity of $\sim$ 0.015 L$\rm_{Edd}$
(\citealt{a67}; although see \citealt{a82}).

\subsection{Column Density : $\rm N_H$}\label{nh_discuss}
Accurate determination of the interstellar column density is crucial due to
the preferential absorption of soft X-rays, which in our case are consistent
with being emitted by the cool accretion disc. Measurement of $\rm N_H$ is
best achieved at soft X-ray energies. Here \textit{XMM} with its well
studied low energy calibration should provide the most reliable
determination of the interstellar column density $\rm N_H = 0.23 \times
10^{22}~cm^{-2}$ \citep{a18}. In contrast the best fit value as measured
with \textit{Suzaku} is $\rm N_H = 0.18 \times 10^{22}~cm^{-2}$. However,
we do note that the measured $\rm N_H$ depends on the continuum model
assumed, e.g. for the Comptonizing corona model we find a best fit value
$\rm N_H = (0.31\pm 0.01) \times 10^{22}~cm^{-2}$. 

In order to investigate the instrumental dependence of our measured values of
the column density, we also fit the best fit models above to a number of
\textit{Swift} XRT spectra taken before and after our \textit{Suzaku}
observation, with the closest occurring $\sim$ 60 days beforehand (obsid:
00030090050). The results from these fits are consistent with the results of
our fits to the \textit{Suzaku} spectrum, with the caveat that the smaller
energy coverage (0.6 -- 10 keV versus 0.6 -- 150 keV) and exposure time
($\sim$ 2 ks) necessarily result in larger confidence regions from fits to
the \textit{Swift} spectra.

These measurements are in agreement with numerous independent measures of
the interstellar column, sourced both from direct measurements and Galactic
surveys. X-ray observations from \textit{Swift} measured a column density of
$\rm 0.20 \times 10^{22}~cm^{-2}$ \citep{a3}, while optical measurements
also require $\rm N_H = 0.2 \times 10^{22}~cm^{-2}$ \citep{a19}. Additionally an
estimate of the column density may be obtained from various radio (N$\rm_H$
= $\rm 0.17 \times 10^{22}~cm^{-2}$ \citealt{a30}, $\rm 0.17 \times
10^{22}~cm^{-2}$ \citealt{a31}), and far-IR ($\rm 0.28 \times
10^{22}~cm^{-2}$ \citealt{a29}) surveys.

The radius of the cool disc component depends on the column density and
inclination as illustrated in Fig. \ref{inner_radius}. From above, we see
that for the directly measured values of $\rm N_H$ towards \swt, the
observed soft excess may originate from an accretion disc whose inner radius
is consistent with the ISCO.

\subsection{A thin-disc at the ISCO?}\label{thin_disc_isco}
The significant soft excess observed in \swt~(Fig. \ref{soft_excess}) is
consistent with an origin in a standard multi-colour blackbody accretion
disc (\citealt{a34} and see Table \ref{specfit_params}). The disc
normalization may be used to estimate the inner radius of the accretion disc
when knowledge of the distance and inclination are available, as norm $\rm
\sim (r_{in}/d_{10~kpc})^2 cos\theta$. This estimate is subject to a number
of corrections, (i) spectral hardening must be accounted for, typically this
requires a multiplicative correction factor $\sim$ 1.7 \citep{a47} and, (ii)
one must correct for the inner radius, where $\rm R_{in} =
1.18r_{in}/\sqrt{cos\theta}$ (\citealt{a61}).  There are also additional
errors, i.e. the zero torque inner boundary condition \citep{a46} and
radiative transfer effects \citep{a35}, which could contribute but are
difficult to quantify.

In Fig. \ref{inner_radius}, we plot the inner disc radius (corrected for (i)
\& (ii) above) measured from our best fit models in \S 3.2. The best fit
radius is consistent with the ISCO, for certain values of $\rm N_H$ \&
inclination, although there are large uncertainties. Previous observations
of \swt~at a 0.5 -- 10 keV flux of 3.9$\rm \times
10^{-10}~erg~s^{-1}~cm^{-2}$ detected a cool accretion disc with an inner
radius consistent with the ISCO. The measurements presented herein, at an
unabsorbed 0.6 -- 10 keV flux of 6.8$\rm \times
10^{-10}~erg~s^{-1}~cm^{-2}$, are in agreement with the previous analysis of
\citet{a18} and furthermore atest to the stability of the cool disc
component given that the observations were separated by almost 18 months
(see Fig. \ref{asmlc}).

\citet{a20} observed GX 339-4 at a luminosity of $\rm \sim 0.05~L_{Edd}$
with \textit{XMM} \& \textit{RXTE}. A significant excess of soft X-ray flux
was detected in the \textit{XMM} data alone, due to the absence of soft
X-ray coverage for \textit{RXTE}. Modelling revealed this excess to be
consistent with a cool accretion disc ($\rm kT_{in} \sim 0.35~keV$)
extending to the ISCO. Subsequent observations utilizing both \textit{RXTE}
\& \textit{Swift} at luminosities of 0.023 and 0.008 $\rm L_{Edd}$ also
detected this excess soft X-ray component \citep{a32}.  Together with the
new observations presented in this paper, these results appear to confirm
the presence of the inner radius of the accretion disc at the ISCO (or at
least at a radius far lower than expected) for luminosities far below that
at which the state transition from HSS to LHS state occurs, in contrast to
theoretical expectations (e.g. \citealt{a36}).

\citet{a74} have re-analysed the \textit{XMM} data on \swt~\& GX 339-4
focusing on the timing characteristics and variability inherent to the
spectra. Using a new analysis technique, they find significant evidence for
the presence of the soft spectral component in agreement with
\citet{a18,a20}. They interpret the soft X-ray flux as originating in
variability intrinsic to the accretion disc. In particular they find that
the intrinsic variability of the accretion disc is likely to be responsible
for the low frequency Lorentzian feature present in the power spectral
density function of sources in the low-hard state. This places a limit on
the disc truncation radius of $<$ 20 $\rm R_g$.  The \textit{XMM} spectrum
of \swt~has also been re-analysed by \citet{a75}. They fit the spectrum with
a relativistically blurred reflection model and find an inner disc radius of
$\rm R_{in} \sim 3~R_g$. This in turn provides an measure of the spin of the
putative black hole of $\rm a \sim 0.76$. Again, inconsistent with the idea
of a large truncation radius for the accretion disc in the low-hard state.

The magnitude of the reflection features is observed to be quite low in our
data. Although the presence of a broad relativistic line is required by the
data ($> 7 \sigma$ significant, EW $\rm \sim 70 \pm 30~eV$), it is
intrinsically weak (norm $\sim 10^{-4}$). The detection of a relativistic
iron line is consistent with the measurements of \citet{a76} who also
measured a redshifted line at an energy of 6.2 keV during pointed
\textit{RXTE} observations (see Fig. \ref{asmlc}). The inner radius inferred
from the {\tt laor} line fits is larger than the ISCO at $\rm R_{in} \sim$
10 -- 20 gravitational radii although much less than might be expected from
the standard disc picture for disc truncation, where the inner radius is
expected to be an order of magnitudes larger \citep{a36}. 

The curvature present in the PIN spectrum is also consistent with a low
reflection fraction, f $\sim$ 0.12 - 0.21 (inclination $\sim$ 30$^{\degr}$
-- 63.256$^{\degr}$). We note that the reflection fraction inferred from the
self consistent {\tt CDID} model is at the lower end of the values inferred
from the phenomenological {\tt reflect*(diskbb+po)+laor} model. It is known
that the reflection features of black hole binaries are generally weaker in
the low-hard state than in higher luminosity states
(i.e. \citealt{a57,a58}).  The relative weakness of these features in
\swt~is primarily due to the low luminosity nature of the source ($\rm L_x
\sim 8 \times 10^{36}~erg~s^{-1}$). In comparison, the X-ray luminosity in
GX 339-4 (\citealt{a20}) was much higher when a strong relativistically
smeared iron line ($\sim 8 \sigma$ significant, EW $\sim$ 350 eV) was
detected ($\rm L_x \sim 4 \times 10^{37}~erg~s^{-1}$).

The value we measure for the reflection fraction in \swt~is comparable to
previous LHS observations where reflection fractions of $\sim$ 0.2 -- 0.3
were measured in Cyg X-1 \& GX 339-4 \citep{a37,a20}. In contrast,
observations at higher luminosities require much higher reflection
fractions. X-ray spectra of the black hole binaries XTE J1650-500 and GX
339-4 in the VHS state revealed reflection fractions of approximately unity
\citep{a57,a58}. The low reflection fractions inferred in the low-hard state
may be interpreted as evidence for a truncated disc, e.g. \citet{a37}.
Alternatively the disc may remain close to the ISCO with the low reflection
fraction resulting from beaming of the hard X-ray flux away from the disc by
a mildly relativistic outflowing corona, e.g. \citet{a38}.

\begin{figure}
\begin{center}
\includegraphics[height=0.33\textheight,angle=-90]{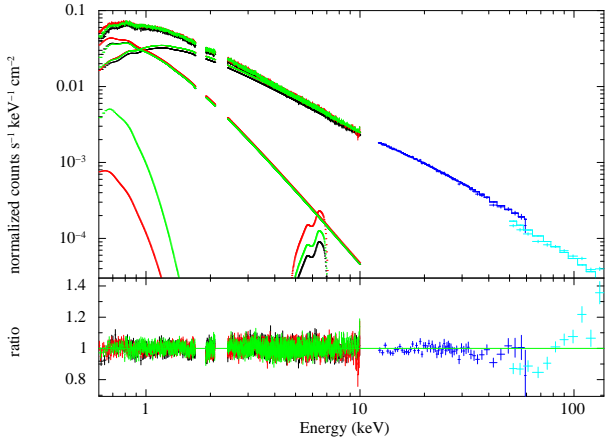}
\caption{The best fit Comptonization model ({\tt
    pha*(diskbb+compps+compps+laor)}) to the \swt~spectrum, see
  \S\ref{spec_ii} for details. The inferred inner disc radius is larger than
  that measured in \S\ref{spec_i}, i.e. $\rm R_{in} \gtrsim 30~R_g$. See
  Table \ref{specfit_params1} for the model parameters.}   
\label{spec_ii_images_2}
\end{center}
\end{figure}

\subsection{Alternatives to a thin disc at the ISCO}
A number of alternative explanations have been put forward to account for
the presence of the soft excess observed in \swt~\& GX 339-4. \citet{a66}
reanalyzed the data from \citet{a18} and find that while it is consistent
with containing a soft disc component, a number of alternative continuum
prescriptions that allow the disc to be truncated are also valid; however,
see \citet{a75,a74} who outline a number of issues with this result.
 
\citet{a64} propose a scenario in which the disc is truncated at larger
radii; however, the inner edge of the disc is irradiated by the high energy
Comptonized photons in the corona. This irradiation will cause the truncated
inner disc edge to be hotter and hence appear to lie at smaller radii than
is actually the case.  In the model of \citet{a65}, the inner edge of the
cool truncated disc is seen to overlap with the inner ADAF region. Here in
the overlap region the cool disc is heated to temperatures mimicking those of
a cool disc at the ISCO.

It is also possible that what we observe is not the accretion disc but
instead a ring of material that has condensed from the corona. Such an idea
has been explored in detail by \citet{a59} and \citet{a60}. They find that
it is possible for enough material to condense from the corona to produce an
inner ring of material extending from near the ISCO to a few tens of
Schwarzschild radii, while the accretion disc is truncated at larger
radii. In particular, detailed modelling was able to reproduce the soft disc
components observed by \citet{a32} in GX 339-4 (see \citealt{a60}).

\subsection{Comparison with previous \textit{Suzaku} observations}
Two additional black hole binaries have been observed by \textit{Suzaku}
while in the low-hard state. \citet{a27} presented observations of GRO
J1655-40, while \citet{a28} observed Cyg X-1. The observations took place
while the systems were in the LHS at luminosities of approximately 0.007
L$\rm_{Edd}$ and 0.02 L$\rm_{Edd}$ respectively.  We also modelled the
spectra of \swt~with a combination of 2 Comptonizing components, to aid
comparison of this data to the \textit{Suzaku} observations of the black
hole binaries Cyg X-1 \citep{a28} \& GRO J1655-40 \citep{a27}. For both of
these systems, the observed spectrum was interpreted as being due to a 2
component Comptonizing corona, i.e. a population of electrons with the same
temperature ($\rm kT_e$) but differing optical depths ($\tau$), modified by
disc reflection, {\tt diskbb+compps+compps}.  The parameters of our best fit
model are listed in Table \ref{specfit_params1}, while the fit itself is
displayed in Fig. \ref{spec_ii_images_2}.  In Table \ref{obs_comparison}, we
display the parameters for our best fit Comptonization model along with
those for the best fit models for Cyg X-1 and GRO J1655-40.

The blackbody disc component is found to be cool (kT $\sim$ 0.1 keV) and the
inner radius of the accretion disc is consistent with the outer radius of
the corona within the errors ($\rm R_{in} \sim 30~R_g$). The best fit radius
for the iron line is also consistent with this value. The high
value for the Hydrogen column density ($\rm 0.31 \times 10^{22}~cm^{-2}$)
returned by this model is inconsistent with the available data and
casts doubt the Comptonization scenario as a viable model for the
observed \textit{Suzaku} spectrum of \swt. However, the presence of
circumstellar matter or a local absorber could account for the excess
absorption above that detected at radio wavelengths.

We find an electron temperature cooler than that measured in both Cyg X-1
and GRO J1655-40 when the spectra are interpreted in terms of this
Comptonizing corona model. Assuming the disc truncation model for the
low-hard state to be correct, we would expect the electron temperature to
decrease with increasing luminosity, e.g. \citet{a73}. The value of $\rm
kT_e$ we measure would imply that \swt~was at a luminosity in excess of that
at which Cyg X-1 was observed ($\rm \sim 0.02~L_{Edd}$). This is in contrast
to various luminosity estimates. Indeed, we measure a luminosity of $\rm
\sim 0.016~d_{8.5kpc}^2M_{10M_{\sun}}~L_{Edd}$; for \swt~to have a
luminosity greater than that of Cyg X-1 at the time of the \textit{Suzaku}
observations would imply a distance greater than 8.5 kpc and or a black hole
mass less than 10$\rm M_{\sun}$.

\citet{a28} have argued that the inner radius of the accretion disc does not
extend to the ISCO in the LHS state based on the \textit{Suzaku}
observations of Cyg X-1 and GRO J1655-40 (see \S 4.3). They base their
argument around 2 main points; firstly the low reflection fractions measured
in the LHS state show that either the accretion disc does not intrude too
deeply into the corona or some form of outflow must be formed. They consider
the outflow case to be unlikely as the reflection fractions measured in both
GRO J1655-40 (i $\rm \sim 70^{\circ}$) and Cyg X-1 (i $\rm \sim 45^{\circ}$)
are similar, whereas if the reflection was due to an outflow it would be
expected to have a strong inclination dependence, which is not
observed. Here, we note that the errors on the reflection fractions measured
in the case of Cyg X-1 (f = 0.4$^{+0.2}_{-0.3}$), call into question the
validity of this inclination based argument, given the quality of the
currently available data.

Secondly it is argued that the inner radius determined from fits to the iron
K line in Cyg X-1 ($\rm \sim 13~R_g$) is inconsistent with the ISCO. This
point is much more ambiguous as their reported value of the inner radius is
$\rm R_{in} = 13^{+6}_{-7}~R_g$. This value is consistent with the ISCO for
a Schwarzschild black hole. As such, it is clear from the currently
available data that there is evidence for a cool disc component, with an
inner radius that in some cases is consistent with the ISCO, whether or not
the cool disc component actually extends to the ISCO is presently not clear.

Finally, it is important to note the high mass X-ray binary nature of Cyg
X-1. Here the accretion process, namely accretion via a stellar wind, is
significantly different from that in GRO J1655-40 (Roche lobe overflow) and
as such may preclude detailed comparison with the low mass X-ray binaries
like \swt~and GRO J1655-40. 

%\begin{landscape} 
\begin{table*}
\begin{center}
\caption{Comptonizing Corona Fits to the Spectra of Low-Hard State Black
  Holes Observed by \textit{Suzaku}.\label{obs_comparison}}
\begin{tabular}{lcccccccccc}
\tableline
System & L$\rm_x$ & N$\rm_H$ & kT$\rm_{bb}$ & R$\rm_{disc}$ & R$\rm_{thick}$ &
R$\rm_{thin}$ & kT$\rm_{e}$ & $\tau_1$ & $\tau_2$ & f \\
       & [ L$\rm_{Edd}$ ] & [ $\rm 10^{21} cm^{-2}$ ] & [ keV ] & [
  km ] & [ km ] & [ km ] & [ keV ] & & & [ $\Omega/2\pi$ ]\\  
\tableline\tableline 
\swt         & 0.016  & 0.31 & 0.1 & $\sim$ 830 & $\sim$ 205 & $\sim$ 360 &
53 &  0.34 & 2.57  & 0 \\ 
GRO J1655-40 & 0.007 & 0.74 & 0.2 & $\sim$ 330 & $\sim$ 26 & $\sim$ 65 & 135 &  0.25 & 1.2  &
0.5 \\ 
Cyg X-1      & 0.02  & 0.66 & 0.3 & $\sim$ 103 & $\sim$ 75 & $\sim$ 200 & 100 &  0.4  & 1.5  &
0.4 \\ 
\tableline
\end{tabular}
\tablecomments{GRO J1655-40 data from \citet{a27}, Cyg X-1 data from
  \citet{a28}, \swt~this work.  Where required a distance of 8.5 kpc and an
  inclination of 63.256$^{\degr}$ have been assumed for \swt. R$\rm_{disc}$
  refers to the inner radius of the accretion disc whereas R$\rm_{thick}$ \&
  R$\rm_{thin}$ refer to the outer radius of the optically thick and thin
  regions of the Comptonizing cloud respectively.}
\end{center}
\vspace{7mm}
\end{table*}
%\end{landscape}

\begin{figure}
\begin{center}
\includegraphics[height=0.25\textheight]{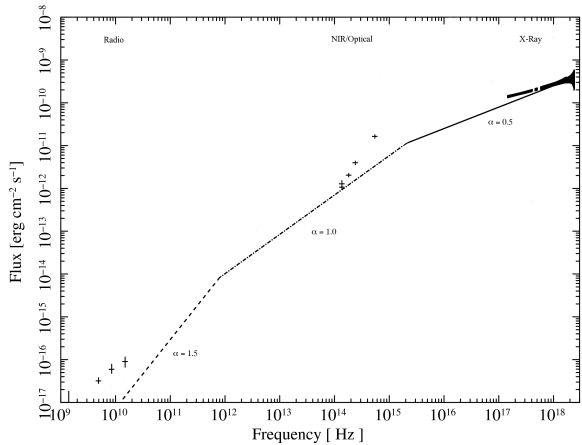}
\caption{The absorption corrected spectral energy distribution for \swt. We
  assumed a column density of $\rm N_H \sim 0.20 \times 10^{22}~cm^{-2}$, to
  facilitate comparison with the SEDs displayed in \citet{a19,a76}. The
  optical/NIR \& X-ray data are quasi-simultaneous, see \S\ref{the_jet}
  for details.}
\label{sed_jet}
\end{center}
\end{figure}

\subsection{Jet and/or Corona?}\label{the_jet}
Our knowledge of role of the relativistic radio jets in X-ray binaries has
advanced rapidly in the past 10 years. Observations of numerous X-ray
binaries have revealed the presence of a compact quasi-steady jet in the
low-hard state, whereas in the high-soft state the jet is typically observed
to disappear \citep{a26,a81}. These also point towards a significant
contribution from the jet towards the overall energy budget of the binary
system, with the jet luminosity being possibly equal to that observed in
X-rays, e.g. \citet{a56}.  In addition, a significant correlation has been
found between the X-ray and radio luminosities of accreting sources across a
range of luminosity (in Eddington units), suggestive of a disc-jet
coupling \citep{a55}.

\citet{a25} have shown how one could, in principle, distinguish between
different jet models based on the magnitude of the reflection features
present in the spectrum. In particular, reflection fractions greater than
0.2 could not be reproduced by their jet models, reflection fractions
between 0.1--0.2 could be due to synchrotron self-Comptonization in the base
of a jet, while a reflection fraction below 10\% could be interpreted as
evidence that the spectrum was dominated by synchrotron emission from the
jet. If the base of the jet is confined within a few gravitational radii of
the black hole, relativistic light bending will become important. This will
act in the opposite direction to the beaming of the outflow leading to a
slightly larger reflection fraction \citep{a80}.  These models are also
consistent with the detection of an accretion disc near the ISCO, as we have
likely detected with \textit{Suzaku}.

A break or cut-off in the high energy spectrum of black holes and neutron
stars can be interpreted in terms of the electron temperature in a thermal
Comptonizing corona. The high energy spectrum of \swt~extends to at least
200~keV without a strong break or cut-off in our observations. Moreover,
single thermal Comptonization models fail to achieve acceptable fits.  In
the context of \citep{a25}, the reflection fraction of $\sim$ 0.12 -- 0.21
measured in \swt~ is consistent with a disc that is illuminated by
synchrotron self-Comptonization produced in the base of a jet. A corona that
is independent of a jet but which produces a non-thermal Comptonization
spectrum might be able to describe the data, but the base of a jet provides
a natural context for the production of hard X-ray emission like that we
have observed.

In section \ref{spec_i}, we showed how the data may be fit with a power-law
modified by reflection, this model is supported by more detailed modelling
in \S\ref{spec_ii}.  In both of these cases an incident power-law is
required, the jet is a natural source of such a power-law. In
Fig. \ref{sed_jet}, we plot the absorption corrected spectral energy
distribution for \swt. The optical data is from \citet{a69} and was taken
during at approximately the same time as our \textit{Suzaku} spectra. In
addition, we include V-band optical and J, H \& K-band NIR data points
obtained in April and July 2009 at the Magellan and MDM observatories. The
V-band magnitude was found to be consistent with the previous measurements
of \citet{a69}, as such the NIR data should be representative of the
infrared portion of the SED at the time of our \textit{Suzaku} observations.
The radio data points represent the radio detection shortly after the 2005
outburst and likely represent upper limits to the radio flux at the time of
our observations \citep{a19}, these are also consistent with the
non-detection at 5 GHz \& 8 GHz by \citet{a77}.

The solid line extending from the X-ray to the optical represents a
power-law as might be expected from optically thin radio emission, i.e.
$\rm f_{\nu} \propto \nu^{-0.5}$.  This is similar to the slope of the
optically thin jet in GX 339-4 \citep{a72}. Extrapolating this power-law to
optical/NIR wavelengths shows that a spectral break is required at
optical/UV frequencies. Analysis of optical data of \swt~in its current
state by \citet{a70} favour a case where any jet emission is insignificant
at optical wavelengths consistent with our extrapolated power-law, this is
in agreement with an independent observation/analysis by \citet{a68,a76}.
\citet{a83} has shown, based on analysis of optical data from XTE J1118+480,
GX 339-4 \& \swt, that the variability at optical wavelengths is
inconsistent with being produced by reprocessing of X-rays in the accretion
disc and instead favour a model where the emission in the low-hard state is
dominated by the jet, e.g. \citet{a84}.

%%%%%%%%%%%%%%%%%%%%%%%%%%%%%%%%%%%%%%%%%%%%%%%%%%%%%%%%%%%%%%%%%%%%%%%%%%%%%
%%%%    Conclusions                                                      %%%%
%%%%%%%%%%%%%%%%%%%%%%%%%%%%%%%%%%%%%%%%%%%%%%%%%%%%%%%%%%%%%%%%%%%%%%%%%%%%%
\section{Conclusions}
We have presented \textit{Suzaku} broadband X-ray observations of the
candidate black hole X-ray binary \swt. The broadband spectrum (2 -- 250
keV) is observed to be consistent with a simple power-law model ($\Gamma
\sim 1.63$). Confirming previous observations, we detect the presence of an
excess at soft X-ray energies. In addition, a weak relativistic iron line
and curvature consistent with a reflection hump at 20 -- 30 keV are
detected.  

These observations point towards the persistence of the accretion disc at a
much lower radius than previously appreciated in the low-hard state.
Estimates of the disc inner radius with both a simple {\tt diskbb+po} model
and a detailed reflection model reveal values consistent with the ISCO ($\rm
R_{in} \lesssim 6~R_g$) for certain values of both the column density and
inclination. In contrast modelling the spectra with a Comptonization model,
while revealing a truncated inner disc ($\rm R_g \sim 50~R_g$), implies a
value for the Hydrogen column density in disagreement with all previous
estimates.

\acknowledgements 
This research has made use of data obtained from the \textit{Suzaku}
satellite, a collaborative mission between the space agencies of Japan
(JAXA) and the USA (NASA). This research has made use of data obtained from
the High Energy Astrophysics Science Archive Research Center (HEASARC),
provided by NASA's Goddard Space Flight Center. This research made 
use of the SIMBAD database, operated at CDS, Strasbourg, France and
NASA's Astrophysics Data System.

GM thanks the Spanish Ministerio de Ciencia e Innovaci\'on and CSIC for support
through a Ram\'on y Cajal contract. J.H. gratefully acknowledges support
from NASA grant NNX08AC20G. 

%%%%%%%%%%%%%%%%%%%%%%%%%%%%%%%%%%%%%%%%%%%%%%%%%%%%%%%%%%%%%%%%%%%%%%%%%%%%%
%%%%    References                                                       %%%%
%%%%%%%%%%%%%%%%%%%%%%%%%%%%%%%%%%%%%%%%%%%%%%%%%%%%%%%%%%%%%%%%%%%%%%%%%%%%%
%     last one was #84

%%%%%%%%%%%%%%%%%%%%%%%%%%%%%%%%%%%%%%%%%%%%%%%%%%%%%%%%%%%%%%%%%%%%%%%%%%%%%
%%%%%%%%%%%%%%%%%%%%%%%%%%%%%%%%%%%%%%%%%%%%%%%%%%%%%%%%%%%%%%%%%%%%%%%%%%%%%
\vspace{1cm}
\footnotesize{This paper was typeset using a \LaTeX\ file prepared by the 
author}

%%%%%%%%%%%%%%%%%%%%%%%%%%%%%%%%%%%%%%%%%%%%%%%%%%%%%%%%%%%%%%%%%%%%%%%%%%%%%
%%%%%%%%%%%%%%%%%%%%%%%%%%%%%%%%%%%%%%%%%%%%%%%%%%%%%%%%%%%%%%%%%%%%%%%%%%%%%

\end{document}